# Superconductivity and non-metallicity induced by doping the topological insulators $Bi_2Se_3$ and $Bi_2Te_3$

Y. S. Hor[1], J. G. Checkelsky[2], D. Qu[2], N. P. Ong[2], and R. J. Cava[1]

[1]Department of Chemistry, Princeton University, New Jersey 08544, USA

[2]Department of Physics, Princeton University, New Jersey 08544, USA

**Abstract**

We show that by Ca-doping the $Bi_2Se_3$ topological insulator, the Fermi level can be fine tuned to fall inside the band gap and therefore suppress the bulk conductivity. Non-metallic $Bi_2Se_3$ crystals are obtained. On the other hand, the $Bi_2Se_3$ topological insulator can also be induced to become a bulk superconductor, with $T_c \sim 3.8$ K, by copper intercalation in the van der Waals gaps between the $Bi_2Se_3$ layers. Likewise, an as-grown crystal of metallic $Bi_2Te_3$ can be turned into a non-metallic crystal by slight variation of the Te content. The $Bi_2Te_3$ topological insulator shows small amounts of superconductivity with $T_c \sim 5.5$ K when reacted with Pd to form materials of the type $Pd_xBi_2Te_3$.

**Introduction**

A novel kind of three-dimensional insulator called topological insulators, which have a bulk band gap but non-trivial topological surface states has been introduced [1–5]. The surface states of these topological insulators show Dirac-like behavior with the spin polarization locked perpendicular to the electron momentum by the effect of strong spin-orbit interaction. Recently, a new class of topological insulators, $Bi_2Se_3$ and $Bi_2Te_3$, previously known to be a good thermoelectric materials [6], has been discovered [7,8,9,10]. The surface states of these materials have been observed by angle-resolved photoemission spectroscopy (ARPES) [7,8,9] and scanning tunneling microscopy (STM) [11,12], but are still considered a challenging problem for electronic transport measurements due to their dominant bulk conductance. Here we show that, by proper chemical doping, we can suppress bulk conduction in $Bi_2Se_3$ and $Bi_2Te_3$, which is necessary for the detection of the surface currents. The electronic transport detection of the surface Dirac state is important for the study of Majorana fermionic physics [13]. As proposed in the reference 13, finding the Majorana fermion requires an interfacing of a topological insulator with a superconductor. We further show that by proper intercalation of transition metal elements, superconductivity can be induced in the $Bi_2Se_3$ and $Bi_2Te_3$ topological insulators.

**Experimental**

The tetradymite $Bi_2Se_3$ and $Bi_2Te_3$ single crystals were grown by a modified Bridgman method, melting stoichiometric mixtures of high purity elements of Bi (99.999 %), Se (99.999 %) or Te (99.999 %) at 800 °C for 16 hours and then slow cooling to 550 °C in sealed vacuum quartz tubes. The crystals were then annealed at 550 °C for three days. Since Ca can react with quartz, a proper procedure is required - the two step method described in ref. 14 was used to grow $Bi_{2-x}Ca_xSe_3$ crystals. To obtain non-metallic $Bi_2Te_3$, the as-grown stoichiometric crystal was annealed under Te vapor at 410 °C for a period of one week with ~0.5 g Te in a sealed quartz tube of volume ~8 cm$^3$. Single crystals of $Cu_yBi_2Se_3$ were grown by melting stoichiometric mixtures of high purity elements Bi (99.999 %), Cu (99.99 %) and Se (99.999 %) at 850 ºC overnight in sealed evacuated quartz tubes. The crystal growth took place via slow cooling from 850 to 620 ºC and then quenching in cold water in order to optimize the quality of the Cu-intercalated $Bi_2Se_3$ crystals. The cleaved silvery surfaces of these crystals turn golden after one day exposure to air, suggesting that exposure to air should be kept to a minimum. $Pd_zBi_2Te_3$ crystals can be grown by the same method; high purity Pd (99.99 %) was used.

All the grown crystals were cleaved very easily along the basal plane, leaving a silvery shining mirror like surface, and were cut into approximately $1.0 \times 1.0 \times 6.0$ mm$^3$ rectangular bar samples for electronic transport measurement. The rhombohedral crystal structure of all the as-grown and annealed crystals was confirmed by x-ray powder diffraction (XRD) using a Bruker D8 diffractometer with Cu Kα radiation and a graphite diffracted beam monochromator. Resistivity and magnetoresistance measurements were performed in a Quantum Design physical property measurement system (PPMS) between 1.8 and 300 K, where the standard four-probe technique with silver paste cured at room temperature was used for the electrical transport contacts on a freshly cleaved surface. The measurements were conducted with the applied magnetic field along the *c*-axis with the electric-currents parallel to the basal plane of the crystals. The potential contacts were made on the cleavage surface. Hall Effect measurements to determine carrier concentrations were performed in a home-built apparatus. AC magnetization measurement was performed in the PPMS with a magnetic field of amplitude 5 Oe.

**Results and Discussions**

*$Bi_2Se_3$*

XRD results confirm the rhombohedral crystal structure of $Bi_2Se_3$, which consists of Bi and Se hexagonal planes stacked along the [001] crystallographic direction or *c*-axis (hexagonal setting), with the atomic order, Se(1)-Bi-Se(2)-Bi-Se(1), where (1) and (2) refer to different lattice positions [15]. The unit cell consists of three of these units stacked on top of each other with weak van-der-Waals bonds between Se(1)-Se(1) layers, making the (001) plane the natural cleavage plane. In reality, an as-grown $Bi_2Se_3$ crystal tends to have Se vacancies, giving rise to an excess of electrons. In reference 14, STM measurements show that Se vacancy defects are present in $Bi_2Se_3$ crystals, giving rise to an excess of electrons in the system which bring the Fermi level into the conduction band. The spectra of $Bi_2Se_3$ obtained from scanning tunneling spectroscopy (*dI/dV*) were *n*-type, as expected, due to the presence of Se vacancies, and was confirmed by ARPES as is schematically sketched in Fig. 1(a). Thus the STM data support the defect chemistry model described above for the Ca-doped $Bi_2Se_3$, with the added observation that the concentration of defects in general is lower in the Ca-doped samples than for native $Bi_2Se_3$ [14].

The temperature-dependent resistivities in the *ab* plane for undoped $Bi_2Se_3$ and $Bi_{2-x}Ca_xSe_3$ crystals with $x = 0.005$ and $0.0025$ are shown in Fig. 1(b). Both $Bi_2Se_3$ and $Bi_{1.995}Ca_{0.005}Se_3$ show the weakly metallic resistivities commonly seen in high carrier concentration small band gap semiconductors, with resistivities in the 0.3 to 1.5 mΩ-cm range at temperatures near 10 K. The lightly doped $x = 0.005$ material is p-type with a carrier concentration at $T = 1.5$ K determined by Hall effect measurements as shown in the inset of Fig. 1(b) to be approximately $1 \times 10^{19}$ cm$^{-3}$, whereas for the n-type Bi$_2$Se$_3$, the carrier concentration is $8 \times 10^{17}$ cm$^{-3}$. For the n-type Bi$_2$Se$_3$ crystal, the valence band and a small pocket of the conduction band near the Γ point are clearly observed below $E_F$ in ARPES data [14] as shown in the left plot of Fig. 1(a). Bi$_{1.995}$Ca$_{0.005}$Se$_3$ has the $E_F$ in the valence band (Fig. 1(a) right) showing a p-type carrier which is consistent with the Hall effect measurement as shown in the inset of Fig. 1(b). This $E_F$ shift indicates that the Ca doped crystal is hole doped relative to the n-type Bi$_2$Se$_3$ crystal. In the p-type doped samples, the Fermi level is below the energy of the Dirac point. By fine tuning with 0.125 % Ca doping in Bi$_2$Se$_3$, we achieve the goal of obtaining a non-metallic Bi$_2$Se$_3$ topological insulator, where its resistivity increases to a very large value of ~70 mΩcm as temperature decreases to 100 K. This is consistent with the ARPES data [16] as sketched in the center of Fig. 1(a) that the $E_F$ is dramatically lowered to lie very close to the Dirac point, and therefore of interest in fundamental and applied studies of the topological surface states [17].

### $Cu_xBi_2Se_3$

Single crystals of Cu-intercalated Bi$_2$Se$_3$, Cu$_y$Bi$_2$Se$_3$ within the composition range $0.10 < y < 0.15$ are reproducibly superconducting whereas single crystals of Cu-substituted Bi$_2$Se$_3$, Bi$_{2-y}$Cu$_y$Se$_3$ prepared for $y = 0$ to 0.15 are never superconducting. For all Cu$_y$Bi$_2$Se$_3$ crystals, we observe from XRD that a subtle increase in the c-axis lattice parameter, from $c \sim 28.67$ Å for Bi$_2$Se$_3$ to $c \sim 28.73$ Å for Cu$_{0.12}$Bi$_2$Se$_3$ [18]. By analogy to the similar behavior in observed related layered Cu$_x$TiSe$_2$ [19], we infer that the intercalated Cu in the van der Waals gap partially occupies the octahedrally coordinated 3b (0, 0, ½) sites in the $R\bar{3}m$ space group [18]. In reference 18, STM has confirmed that the Cu is indeed intercalated in between $Bi_2Se_3$ layers and also showed that there is no Cu cluster formation within the van der Waals layers below the exposed surface.

Figure 2(a) shows the temperature dependence of the resistivity of the $Bi_2Se_3$, $Cu_{0.1}Bi_2Se_3$, $Cu_{0.12}Bi_2Se_3$, and $Cu_{0.15}Bi_2Se_3$ single crystals, measured in the *ab* plane. The resistivities of the crystals show weakly metallic in the whole range of temperature except for the $Bi_2Se_3$ that has slight increase in its resistivity at temperatures below 20 K. The $Cu_yBi_2Se_3$ superconducting crystals have about one order magnitude lower resistivity than $Bi_2Se_3$ which is consistent with the Hall effect measurement [18]. The Hall effect measurements indicated that these superconducting crystals are *n*-type, with a temperature independent carrier density of approximately ~ $2 \times 10^{20}$ $cm^{-3}$. This carrier concentration is one order of magnitude higher than is found in native $Bi_2Se_3$ [20,21], and two orders of magnitude higher than is found for crystals with chemical potentials tuned by Ca doping [14,17]. The superconducting transition for $Cu_{0.12}Bi_2Se_3$, which is expanded in the inset of Fig. 2(a), occurs at ~ 3.8 K. The resistivity does not drop to zero below $T_c$ however, indicating that there is not a continuous superconducting path in the crystal. We attribute this to the sensitivity of the superconducting phase to processing and stoichiometry as discussed in reference 18.

Zero field cooled (ZFC) ac magnetization of the single crystals of $Cu_yBi_2Se_3$ for *y* = 0.1, 0.12, and 0.15 is shown in Fig. 2(b). Independent of *y*, the superconducting transition temperature is approximately 3.8 K, but the signal amplitude is increased with the increase of *y*, giving the highest amplitude as ~0.025 emug$^{-1}$ at 1.8 K for *y* = 0.15.

This amplitude is about 20 % of that expected for full diamagnetism, but that estimate represents a lower limit because temperatures substantially below the transition were not accessible in our apparatus and the transition is not complete at the temperature where the field is applied for the ZFC measurement. The inset of Fig. 2(b) shows field dependence of the transverse resistivity, $\rho_{xx}$ for $Cu_{0.12}Bi_2Se_3$ crystal with the magnetic field applied parallel to the *c*-axis. The upper critical field $H_{c2}$ can be determined, confirming that the system is a type II superconductor.

### *$Bi_2Te_3$*

$Bi_2Te_3$ has identical structure to $Bi_2Se_3$. Similar to the $Bi_2Se_3$ case, as-grown $Bi_2Te_3$ always shows metallic behavior. In contrast to $Bi_2Se_3$, stoichiometric $Bi_2Te_3$ crystals tend to have anti-structure defects in which Bi atoms go on a Te site [22], resulting *p*-type material with hole concentration near $10^{19}$ cm$^{-3}$. Due to the complexity of the Bi-Te thermodynamic phase diagram [23], even the most perfect crystals have a relatively large number of antisite defects. This in fact leads to variations in the electronic properties of as-grown crystals. Unlike $Bi_2Se_3$, where acceptors such as Ca can be introduced into the system to compensate the defect caused by Se vacancies, the antisite defect in $Bi_2Te_3$ can be hardly resolved. However, by annealing $Bi_2Te_3$ crystals in closed systems with polycrystalline buffer material just below the melting point, the defect concentrations can be varied [23]. Here, we employ lower temperature annealing of initially *p*-type crystals in Te vapor to obtain non-metallic samples. Fine tuning with this technique to bring the $E_F$ into the bulk gap is possible in the $Bi_2Te_3$ system.

The electronic properties of $Bi_2Te_3$ can be changed by slight variation of the Te content in the crystal. Figure 3 shows the transport property as a function of temperature for *S1*, *S2*, *S3*, and *S4* samples. Most of the as-grown crystals, just like *S1*, exhibit metallic behavior over the whole range of temperatures. However, non-metallic behavior can be achieved through a slight increase of the Te content in the crystal by Te-vapor annealing. Non-metallic samples, *S2* and *S3* show a very different temperature dependence of resistivity. For *S3*, it behaves like a metal on decreasing temperature from 300 to 180 K, where a minimum is reached. As temperature decreases, the resistivity increases, reaching the maximum of 2.7 mΩcm at $T \sim 100$ K, where it then reverts to metallic behavior down to ~20 K and finally rises slightly below 20 K. The upper inset of Fig. 3 shows the sketch of the surface state dispersion [9] near the Γ point where the dash lines indicate the suggested Fermi levels of the *S1*, *S2*, and *S3* samples.

The lower panel of Fig. 3 shows the influence of magnetic field on the temperature dependent transport properties of the *S4* crystal, which is found to exhibit a significant magneto response in the entire temperature range below 300 K. In a 9 T field, it has magnetoresistance (defined as $MR = \frac{\rho_H - \rho_0}{\rho_0} \times 100\%$ where $\rho_H$ and $\rho_0$ are the resistivities at *H* and 0 applied magnetic fields respectively), *MR* ~1400 % at 2 K and ~30 % at room temperature. The temperature dependent resistivity becomes more insulating in 9 T applied field. The lower inset in Fig.3 shows the temperature dependent Hall densities of the *S4* crystal. Hall resistivity measurements performed at magnetic fields close to 9 T, yield the Hall densities of the *S4* sample as $1.7 \times 10^{19}$ cm$^{-3}$ at room temperature and $9 \times 10^{18}$ cm$^{-3}$ at below 20 K.

***Pd$_z$Bi$_2$Te$_3$***

Superconductivity can be found in Pd-added $Bi_2Te_3$, with a superconducting transition at about 5.5 K. The as-grown $Pd_zBi_2Te_3$ crystals for $z$ = 0.1, 0.15, 0.3, 0.5, and 1 have an identical structure to $Bi_2Te_3$ crystal but with a slight change in $c$ parameter. XRD shows that there are minor impurity phases, such as $PdTe_2$, BiPdTe, and $Pd_{72}Bi_{19}Te_9$ present in some samples, but none of them shows a superconducting transition at 5.5 K. However, BiPdTe, which is a known superconductor with the reported $T_c$ as 1.2 K [24], shows a superconducting transition with the onset $T_c$~2 K in our magnetic susceptibility and resistivity measurements.

In Figure 4, we show the ZFC ac magnetic susceptibility data for the Pd-added $Bi_2Te_3$ crystals. $Pd_zBi_2Te_3$ crystals with $z$ = 0.15, 0.3, 0.5, and 1 show the superconducting transition at about 5.5 K, whereas $Bi_{2-x}Pd_xTe_3$ and other impurity phases do not. Besides, there is a 2 K superconducting transition present in some crystals, which is attributed to the presence of the BiPdTe phase. These transitions are consistently observed in the resistivity data, which is plotted in the inset of Figure 4 as the normalized resistivity with respect to the 10 K resistivity, $\rho/\rho_{10K}$ for $z$ = 0.3, 0.5, and 1, with the measured 10 K resistivities being 96, 92, and 146 μΩcm respectively. Though the magnetic susceptibility plots of Pd-added $Bi_2Te_3$ superconductors have a broader transition below 5.5 K, the resistivity plots have a sharp transition. However, the resistivities do not drop to zero below the transition, which could be due to the same situation as present in the $Cu_yBi_2Se_3$ superconductors. The superconducting phase fraction is small in $Pd_xBi_2Te_3$, which is estimated to be about 1 % of the superconducting volume fraction, based on the ac magnetic susceptibility signal at 2 K for $z$ = 1 compound.

We suggest that the data show that Pd intercalated $Bi_2Te_3$ gives rise to superconductivity, just like Pd intercalated $TiSe_2$ [25] shows a superconductor with $T_c \sim 2$ K. The small fraction of superconducting volume in the $Pd_zBi_2Te_3$ crystals could be due to the difficulty in having the Pd intercalated in between the $Bi_2Te_3$ layers. Besides, the antisite defects occurring in $Bi_2Te_3$ may prevent the intercalation. Whether this is true or not, more work is needed in the future, such as STM and TEM measurements, to confirm this argument.

**Conclusion**

We have shown that one can tune the electronic properties of $Bi_2X_3$ topological insulators by doping effects. Through consideration of the defect chemistry of $Bi_2Se_3$, we have identified Ca as a dopant that when present in small percent quantities results in the formation of *p*-type charge carriers and further decreases its bulk conductivity. Ca-doping to $Bi_{1.9975}Ca_{0.0025}Se_3$ allows the Fermi level to be tuned to the Dirac point, making this material of substantial interest for the physics and technological studies proposed for topological insulators. *p*-type $Bi_2Te_3$ can also be tuned into an insulator by annealing the crystal at Te vapor at a low temperature. Use of this simple annealing method facilitates the tuning of the carrier concentrations in $Bi_2Te_3$ crystals to allow for study of surface transport associated with the topological surface states. Besides, these topological insulators can be doped to display superconductivity at accessible temperatures. We have shown that the topological insulator $Bi_2Se_3$ ($Bi_2Te_3$) can be made into a superconductor by Cu (Pd) intercalation between the $Bi_2Se_3$ ($Bi_2Te_3$) layers. This implies that Cooper pairing in the topological insulators is possible. Because of their structural compatibility, devices with interfaces between a $Bi_2Se_3$ ($Bi_2Te_3$) topological insulator and $Cu_yBi_2Se_3$ ($Pd_zBi_2Te_3$) superconductor can be fabricated for investigating novel concepts in physics based on topological surface states.

## Acknowledgements

This work was supported by the NSF MRSEC program, grant DMR-0819860.

**Figure Captions**

**Fig. 1.** (a) Tuning the bulk Fermi level, $E_F$ in Ca-doped $Bi_2Se_3$. ARPES images in ref. [Hsieh nature] showed that the $E_F$ lies in the bulk conduction (BC) band for $Bi_2Se_3$ (left), inside the bulk band gap for $Bi_{1.9975}Ca_{0.0025}Se_3$ (center), and in the bulk valence band (BV) for $Bi_{1.995}Ca_{0.005}Se_3$ (right). (b) Temperature dependent resistivity in the *ab* plane of $Bi_2Se_3$, $Bi_{1.9975}Ca_{0.0025}Se_3$, and $Bi_{1.995}Ca_{0.005}Se_3$ crystals. The stoichiometric $Bi_2Se_3$ crystal is *n*-type and the Ca-doped materials are all *p*-type. The inset shows the Hall effect data at $T=1.5$ K [14] employed to determine the carrier concentrations for a $Bi_2Se_3$ crystal with an electron carrier concentration of $8 \times 10^{17}$ cm$^{-3}$ and $Bi_{1.995}Ca_{0.005}Se_3$ crystal with a hole carrier concentration of $1 \times 10^{19}$ cm$^{-3}$.

**Fig. 2.** (a) Temperature dependent resistivity of $Bi_2Se_3$, $Cu_{0.1}Bi_2Se_3$, $Cu_{0.12}Bi_2Se_3$, and $Cu_{0.15}Bi_2Se_3$ single crystals with applied current in the *ab*-plane. The inset shows the resistivity behavior of a $Cu_{0.12}Bi_2Se_3$ crystal for $1 < T < 6$ K. (b) The temperature dependent magnetization of $Cu_{0.1}Bi_2Se_3$, $Cu_{0.12}Bi_2Se_3$, and $Cu_{0.15}Bi_2Se_3$ single crystals shows a superconducting transition $T_c \sim 3.8$ K in zero field cooled (ZFC) ac susceptibility measurements. The inset shows field dependence of the transverse resistivity, $\rho_{xx}$ at several temperatures below 5 K for a single crystal of $Cu_{0.12}Bi_2Se_3$ with the magnetic field applied parallel to the *c*-axis.

**Fig. 3.** The upper panel shows the resistivities as a function of temperature of the as-grown and annealed $Bi_2Te_3$ crystals labeled as *S1*, *S2* and *S3*. The inset in the upper panel shows the sketch of the surface state dispersion [9] near the Γ point where the dash lines indicate the suggested Fermi levels of *S1*, *S2*, and *S3* samples. The lower panel shows the temperature dependent resistivity of *S4* $Bi_2Te_3$ crystal in zero and 9 T applied magnetic

fields. Hall density $n_H$ of *S4* $Bi_2Te_3$ crystal is plotted in the lower inset for the whole range of temperatures.

**Fig. 4.** Temperature-dependent ac magnetic susceptibilities for $Pd_zBi_2Te_3$ crystals, with $z$ = 0.15, 0.3, 0.5 and 1. No superconductivity is found in $Bi_{2-z}Pd_zTe_3$, $PdTe_2$, and $Pd_{72}Bi_{19}Te_9$. The inset depicts the low-temperature resistivity data for $Pd_zBi_2Te_3$, showing the superconducting transition $T_c \sim 5.5$ K for $z$ = 0.3, 0.5 and 1.

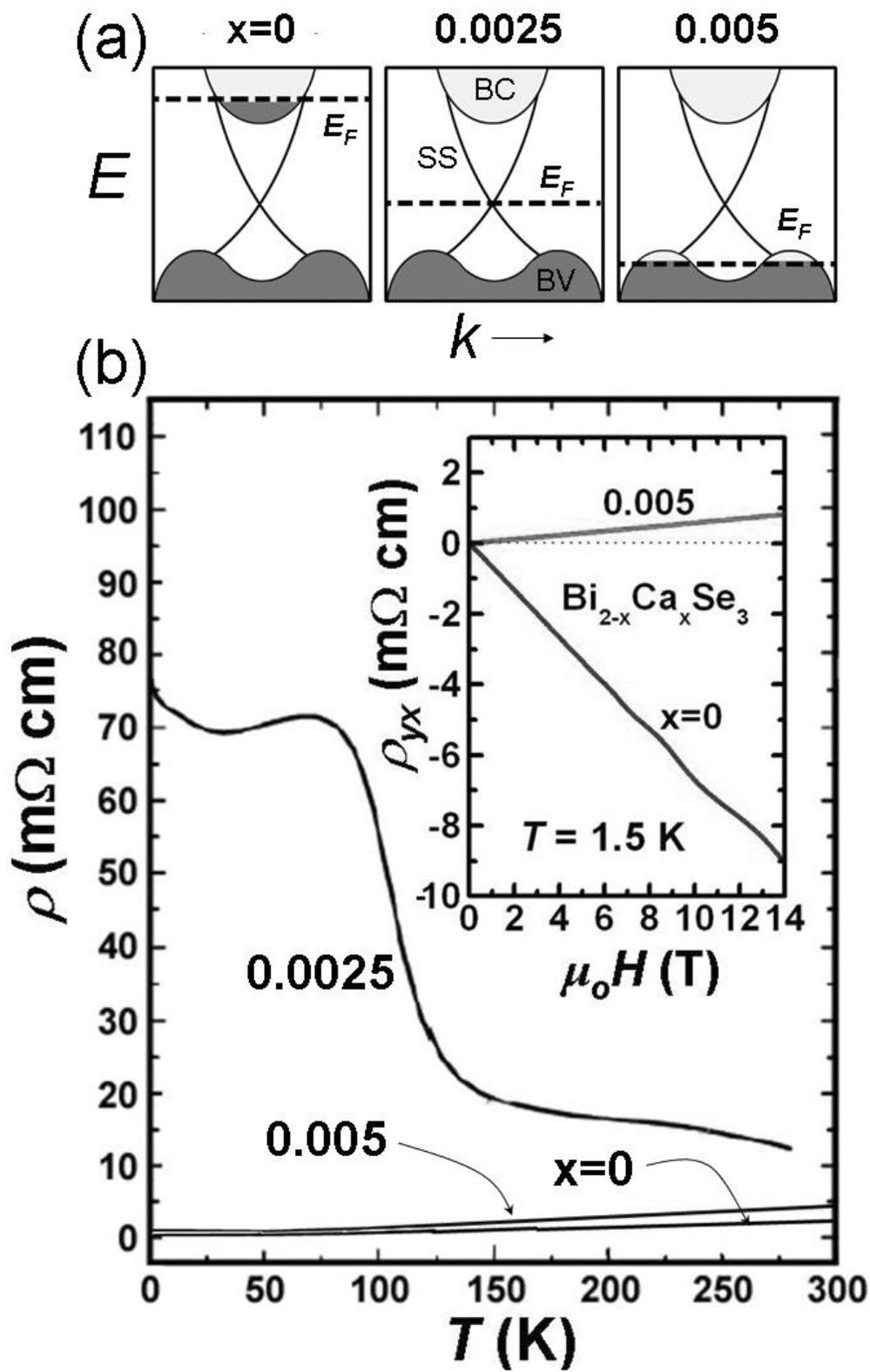

Figure 1

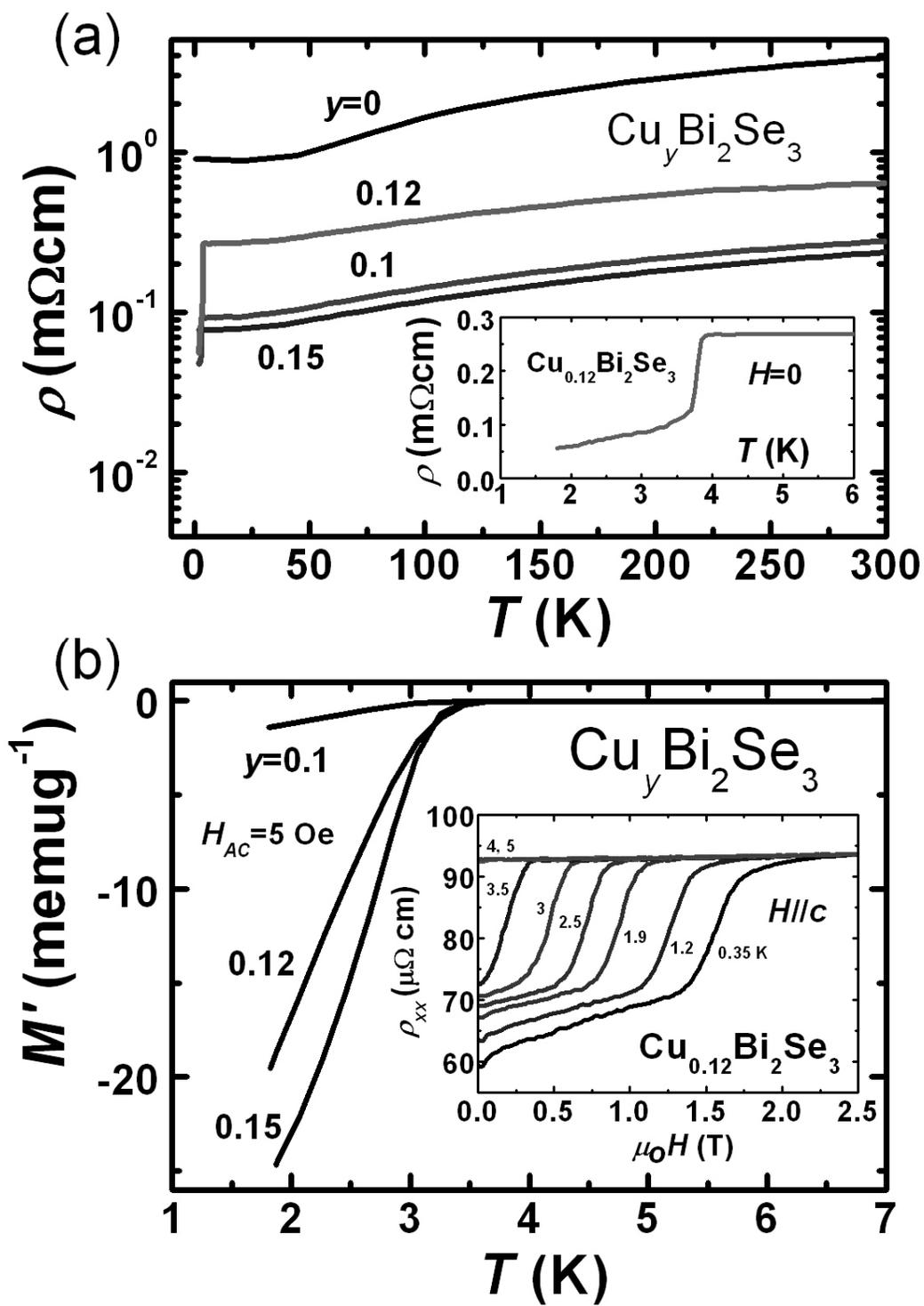

Figure 2

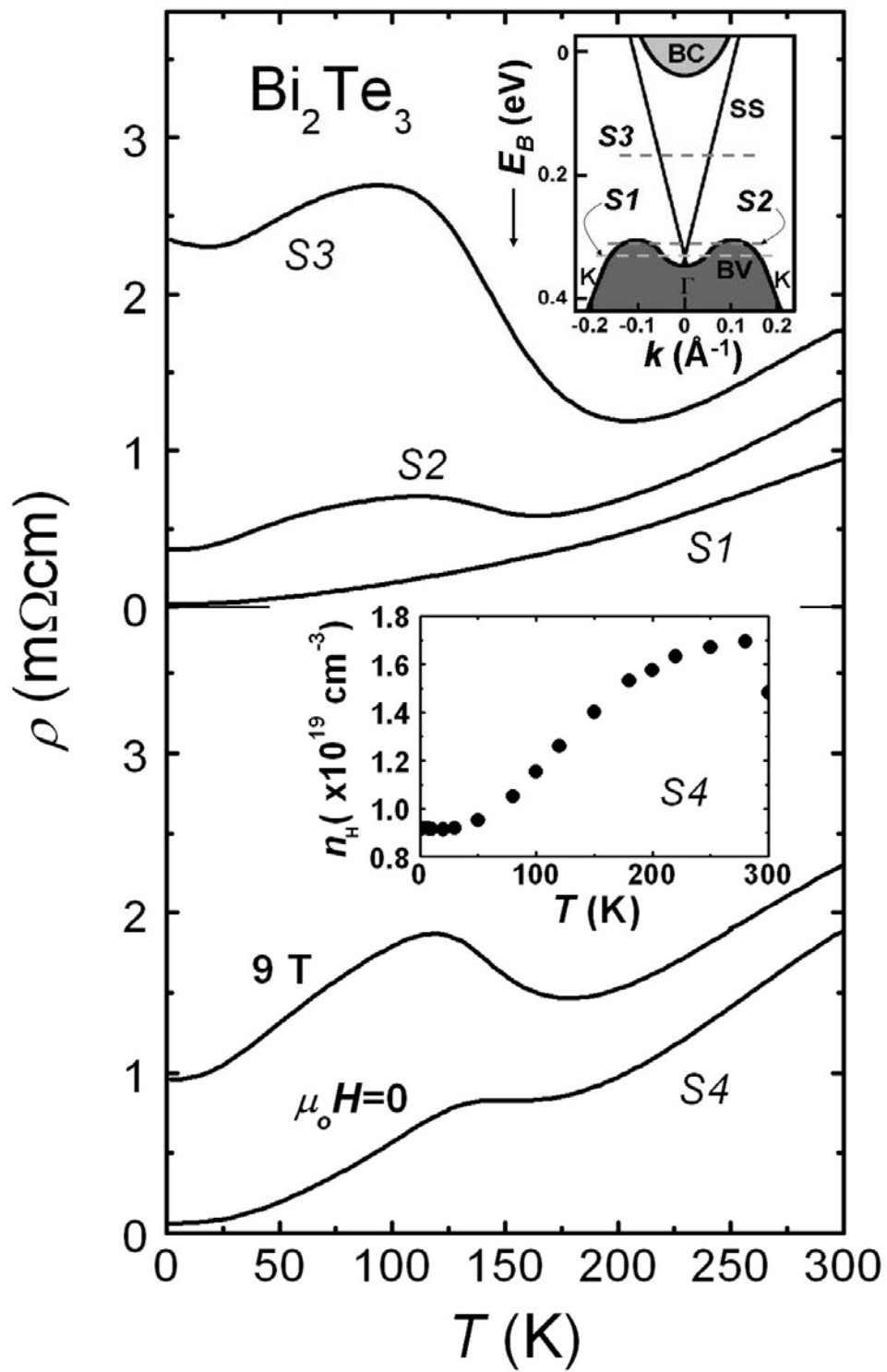

Figure 3

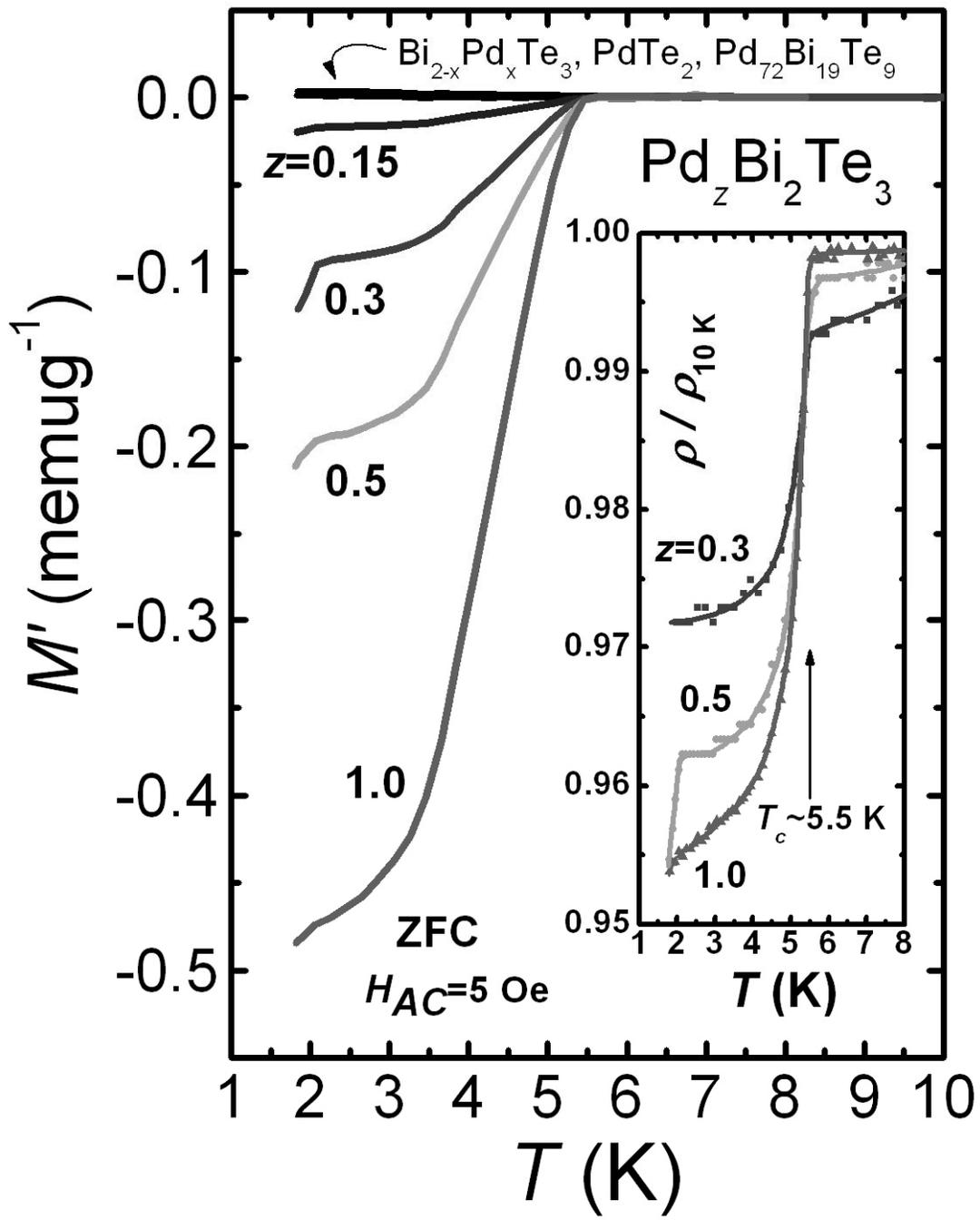

Figure 4